\documentclass[twocolumn,showpacs]{revtex4}
\usepackage{graphicx,subfigure,color}
\usepackage{bm,xspace}
\makeatletter

\newcommand{\rev}[1]{{\color{black}#1}}
\begin{document}

\def\bra#1{\mathinner{\langle{#1}|}}
\def\ket#1{\mathinner{|{#1}\rangle}}
\def\braket#1{\mathinner{\langle{#1}\rangle}}

\preprint{This line only printed with preprint option}
\title{Controlled cavity-QED using a
photonic crystal waveguide-cavity system}
\author{Peijun Yao}
\author{S. Hughes}
\email{shughes@physics.queensu.ca}
\address{Department of Physics, Queen's University\\
Kingston, ON  K7L 3N6 Canada}
\begin{abstract}


\rev{We  introduce a
photonic crystal waveguide-cavity system
for controlling single photon cavity-QED processes.
%
Exploiting Bloch mode analysis, and
medium-dependent Green function techniques,
we demonstrate that the propagation of single photons can be
accurately  described analytically, for integrated periodic waveguides with little more than
four unit cells, including an output coupler.
We verify our analytical approach by comparing to rigorous numerical calculations
for a range of photonic crystal waveguide lengths.
This allows one to
nano-engineer various regimes of cavity-QED with unprecedented control.
We demonstrate Purcell factors of greater than 1000 and  on-chip single
photon beta factors of about 80\% efficiency. Both weak and
strong coupling regimes are investigated, and the important role
of waveguide length on the output emission spectra is shown,
for vertically emitted emission and output waveguide emission.}
\end{abstract}

\pacs{42.50.Pq, 41.20.Jb,  42.70.Qs}

\maketitle

\rev{
\section{Introduction}}

Single semiconductor quantum dots (QDs) are
promising candidates for single photon emission
 applications because  of their unique attributes, e.g.,
 large exciton dipole moments,  integrability with compact
semiconductor cavity systems~\cite{moreau, santori2, pelton, hennessy, englund}, and
compatibility with telecom components.
They also facilitate the study of  light-matter interaction
at a very fundamental level.
However, semiconductor QDs suffer from
environment-induced decoherence~\cite{kuhnPRB02}, that can have
a
detrimental influence on the desired ``indistinguishable'' and coherent nature of the emitted photons. In the last few years, there have been a number of experiments  that show that these shortcomings can be largely overcome by increasing the spontaneous emission rate due to the Purcell effect~\cite{purcell}, which is achieved by coupling the QD exciton to a target cavity mode. For example, planar photonic crystal (PC) cavities, such as those pioneered  by Akane {\em et al.}~\cite{noda},  allow a pronounced modification of the single photon decay,
by careful spatial and spectral positioning of an embedded
QD exciton~\cite{badolato}.

While new regimes of semiconductor cavity-QED (quantum electrodynamics) are
being experimentally realized using photonic nanocavities,
one major drawback
of the  monolithic cavity  is that the
photons
are typically
emitted out of the cavity and thus cannot be efficiently collected
and manipulated. Moreover, it is against the general vision of
{\em planar integration}, as one ultimately wants to emit the
photons on-chip, into a target propagating mode;
compared to
regular microcavity systems, PC
waveguides have the inherent advantage that they can collect and control the  photons
on-chip~\cite{thorhauge,fan,wu,jelOL2007,
YaoPRL2005, GaoAPL2008, YangPRL2009}. Moreover, enhanced spontaneous emission does not even need a
quasi-closed cavity, and {\em open system
cavity-QED} can be exploited to achieve photon emission enhancements by
appropriate bandgap engineering of the propagation modes~\cite{hughes2004,hughesprb,lecamp}.
Related experiments on PC waveguides have  been performed
by Viasnoff-Schwoob {\em et al.}~\cite{viasnoffschwoob} and by
Lund-Hansen {\em et al.}~\cite{lund}; though only modest
Purcell factors were achieved, the waveguide
results of Ref.~\cite{lund} demonstrated that large beta
factors can be achieved for emission into an on-chip waveguide mode.
However, several 
problems remain with long waveguide samples:
 since slow waveguide modes are required to increase
the local density of states (LDOS), then large disorder-induced
propagation losses occur~\cite{Hughes:2005,Povinelli:2004,Gerace:2004}
and  the LDOS peak largely broadens~\cite{Fussell:2008};
in addition,
for on-chip applications, one needs efficient output
coupling, which requires a coupler and an output
(non-PC) waveguide.
Improvements for single photon
gun applications have been proposed~\cite{manga_photongun}
using a small section  of a PC waveguide that mimics a slow-light mode;
although
improved single photon applications were demonstrated,
drawbacks of the finite-size PC waveguide include: $i$) longer waveguides
are required to obtain large Purcell factors ($>100$), and observing the
strong coupling regime would be difficult; $ii$) lack of tunability
and separation of the QD coupling region with the output coupling region;
$iii$) complex Fabry P\'erot ripples appear on the LDOS profile which can be challenging
to overcome and engineer; $iv$) lack of theoretical insight using the known
modes of the system, thus requiring a complex 3D numerical solution where parameter
design sweeps are not practical; $v$) last,  the waveguide
 looses  many of the benefits of a PC nanocavity, e.g., local tuning and pronounced QD coupling using
best-of-breed $Q/V_{\rm eff}$
ratios, where $Q$ is the quality factor
and $V_{\rm eff}$ is the effective mode volume.

In this work, we introduce a  hybrid solution for controlled cavity-QED, that combines
 the benefits of finite-size waveguides, on-chip couplers, and
PC nanocavities, integrated together on a PC planar chip. Although
a rather complicated structure to model and understand,
we show that
Bloch mode analysis and Green function theory can be applied
to present a quantitative solution to the full
scattering geometry. Our \rev{medium-dependent} quantum
optics theory
is supported by
numerically exact solutions of the
3D Maxwell equations.
A schematic of the proposed device is shown in
Fig.~\ref{fig:schematic}.
Similar integrated devices have been built and measured,
and we adopt, and optimize, the coupler design of Banaee {\em et al.}~\cite{Banaee:APL2007}.
To facilitate single photon emission, an excitation laser
can either excite the QD coherently or incoherently.
Once excited, the QD exciton will couple
to vacuum fluctuations and emit a photon. In the presence
of the PC system, this coupling can be controlled, in such
a way as to, e.g., maximize the probability of photon emission
to the left output channel of the on-chip
waveguide.\\


 \begin{figure}[t]
\centering
\vspace{1mm}
{\includegraphics[width=0.44\textwidth,angle=0]{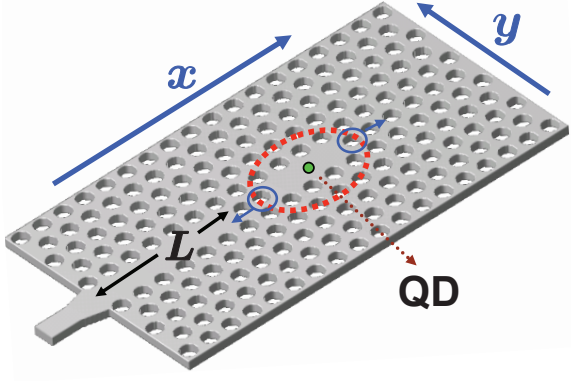}}
 \caption{\label{fig:schematic}(color online) Schematic diagram of the waveguide-cavity single photon source, which is composed of one cavity, one waveguide and  a QD (indicated by green filled circle, which would nominally be located at  slab  center). The PC waveguide length is $L$. The blue circled holes are shifted outwards to increase
 the cavity quality factor.}
\end{figure}

 \rev{
\section{Theory}
\subsection{Medium Green Functions}
{\em Photonic crystal waveguide plus output waveguide.--}
We first derive the Green function of PC waveguide coupled to a semi-infinite output waveguide, as shown in Fig.\,2, but
{\em excluding} the cavity. The PC waveguide has a finite size $L$, and the reflection coefficient is one
 (perfect PC without the cavity) at the right, and $r$ at the left. The electric-field eigenmode of this structure in
the PC waveguide space ($0<x<x_0$) is
 \begin{eqnarray}
{\bf f}_k(\bm r)\!=\!\frac{\sqrt{a/L_n}}{1-re^{2ikL_{\rm eff}}}[{\bf e}_k(\bm r)e^{ik(x-x_0)}+
{\bf e}_{-k}(\bm r)e^{-ik(x-x_0)}] , \
\end{eqnarray}
where $a$ is the pitch, $L_n\rightarrow \infty$ is the normalized length of
the infinite PC waveguide, with eigenmode $e_k(\bm r)$, and
 $L_{\rm eff}$ is an {\em effective} optical length that, for the calculations below,
is found to be $L_{\rm eff} \approx L + 0.38a$~\cite{phase}.
The Green function
  is defined from
\begin{eqnarray}
	\left[\nabla \times \nabla \times  - \frac{\omega^2}{c^2} \varepsilon({\bf r})\right] {\bf G}({\bf r},{\bf r};\omega)
	= \frac{\omega^2}{c^2} \mathbf {1} \delta({\bf r} - {\bf r}'),
\end{eqnarray}
where ${\mathbf 1}$ is the unit dyadic and $\varepsilon({\bf r})$ is the dielectric constant for the material,
and ${\bf G} ={\bf G}^{\rm T} + {\bf G}^{\rm L}$ includes both transverse and longitudinal
contributions.
 The
one-end-closed  waveguide Green function  can be expressed as
\begin{eqnarray}
{\bf G}^{\rm T}_w({\bf r},{\bf r'};\omega) =\sum_k \frac{\omega^2}
{\omega_k^2-\omega^2}\mathbf f_k({\bf r}) {\mathbf
f}_{k}^*({\mathbf r'}),
\end{eqnarray}
where ${\bf G}^{\rm T}_w$ is the transverse Green function. Without loss of generality, we assume $k>0$.
Replacing the $k-$summation ($k \equiv k_x$) by an integral,
i.e. $\sum_k\rightarrow
 \int_{k=0}^{\infty}\frac{L_n}{2\pi}dk$, then
 \begin{figure}[t]
\centering
\vspace{1mm}
{\includegraphics[width=0.46\textwidth,angle=0]{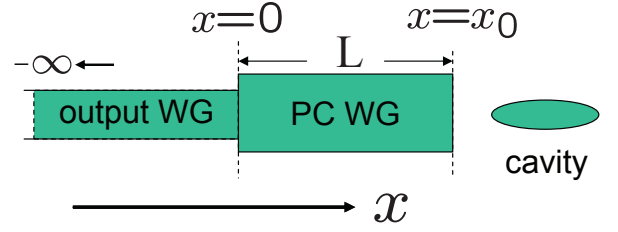}}
 \caption{\label{fig:schematic}(color online) Simple component diagram of the waveguide-cavity system to aid
 the description
of the theoretical formalism; the components include  one cavity, one finite-size PC waveguide and  an infinite (or sufficiently long) output target waveguide at the left ($x<0$).}
%
\end{figure}
 \begin{eqnarray}
 \label{eq:green}
\!\!\!\!{\bf G}^{\rm T}_w({\bf r},{\bf r'};\omega)\!\!
&=&\!\!\frac{L_n}{4\pi}\int_{0}^{\infty}dk\frac{\omega}{\omega_k-\omega-i\delta}\mathbf
f_k({\bf r}) {\mathbf f}_{k}^*({\mathbf r}'), \nonumber\\ \ \ \ \
\!\!&=&\!\!\frac{L_n}{4\pi v_g}\int_{0}^{\infty}dk\frac{\omega}{k-k_{\omega}-i\delta}\mathbf
f_k({\bf r}) {\mathbf f}_{k}^*({\mathbf r}'),
 \end{eqnarray}
where the group velocity $v_g(\omega)$ is treated  as positive and $\delta$ is a positive infinitesimal variable.
Substituting ${\bf f}_k$ from Eq.~(1), and carrying out the complex integration,
\begin{eqnarray}
{\bf G}^{\rm T}_w({\bf r},{\bf r'};\omega)\left|_{0<x<x_0 \atop 0<x'<x_0}
\!\!\!\!\right.&=&\!\!\frac{ia\omega}{2v_g(1-2r\cos(2k_\omega L_{\rm eff})+r^2)}\nonumber\\&&[\Theta(x-x')\mathbf{e}_{k_\omega}(\mathbf
r)\mathbf{e}_{k_\omega}^*(\mathbf
r')e^{ik_\omega(x-x')}\nonumber\\
&+&\Theta(x'-x)\mathbf{e}_{k_\omega}^*(\mathbf
r)\mathbf{e}_{k_\omega}(\mathbf
r')e^{-ik_\omega(x-x')}\nonumber\\
&+& \mathbf{e}_{k_\omega}^*(\mathbf
r)\mathbf{e}^*_{k_\omega}(\mathbf r')e^{-ik_\omega(x+x'-2x_0)}],
\end{eqnarray}
and
\begin{eqnarray}
&&{\bf G}^{\rm T}_w({\bf r},{\bf r'};\omega)|_{x\ll 0 \atop 0<x'<x_0}
=\frac{iat\omega[\mathbf{f}_{k_\omega}^{\rm o}(y,z)]^* e^{-ik_\omega^{\rm o} x}}{2v_g(1-2r\cos(2k\omega L_{\rm eff})+r^2)}\nonumber\\&&[\mathbf{e}_{k_\omega}(\mathbf
r')e^{ik_\omega x'}+
\mathbf{e}^*_{k_\omega}(\mathbf r')e^{ik_\omega(2x_0-x')}],
\end{eqnarray}
where  $\Theta(x-x')$ is the
Heaviside function, 
and $t$ is the transmission amplitude of the PC waveguide mode
 into the output waveguide, which has a propagating mode  ${\bf f}^{\rm o}_{k_\omega}(x,z) e^{i k_\omega^{\rm o}}$, normalized
through $\int_{-\infty}^{\infty}\int_{-\infty}^{\infty} dy dz\, \varepsilon(y,z)\,|{\bf f}^{\rm o}_{k_\omega^{\rm o}}(x,z)|^2 =1 $. In deriving the above equations,
we are assuming that
the PC waveguide has enough unit cells that a Bloch mode description is valid, and that 
the waveguide mode is below the light line with a frequency within the photonic band gap; later,
we will quantify these assumptions with rigorous numerical calculations.

{\em Adding in the cavity.--}
Next, we add a finite single-mode cavity to the finite waveguide plus output waveguide system (see Fig.~2). 
As above, all frequencies are assumed to be deep inside the in-plane photonic band gap. 
The eigenmode of the cavity
${\mathbf f}_c$ with a resonance frequency of $\omega_c$.
Note that these values are for the cavity system
shown, including perfectly matched layers  at the PC waveguide interface (and thus no scattering back
from the $x=x_0$ interface);
the presence of this interface causes a resonance shift and broadening in comparison
to the infinite PC bare cavity eigenmode ${\mathbf f}_c$, e.g.,
 a cavity surrounded by an infinite PC.
For this waveguide-cavity system, we can obtain the
photon Green function following a similar approach of Cowan
and Young~\cite{young04}, and
Hughes and Kamada~\cite{shughesprb04}.
Specifically, we expand the transverse Green function ${\mathbf G}_{wc}^{\rm T}$ of the
PC waveguide-cavity  system in terms of  the cavity and waveguide
eigenmodes, $\mathbf G_{wc}^{\rm
T}=\sum_{\alpha,\beta} B_{\alpha\beta}\, {\mathbf f_\alpha}\otimes
{\mathbf f_\beta}^*$,
 where  ${\mathbf f_{\alpha/\beta}}$ are the
transverse eigenmodes of the uncoupled (separate) waveguide and cavity.
From the definition of ${\bf G}_{wc}^{\rm T}$, we then obtain a set of
equations in matrix form: $MBT=T$. The  matrix
$\mathbf{M}$ has the form
\begin{displaymath}
\mathbf{M} =\left( \begin{array}{ccc}
M_{cc} & M_{ck} &\ldots \\
M_{kc} & M_{kk} & \ldots \\
\vdots & \vdots& \ddots\end{array} \right) ,
\end{displaymath}
with $M_{11}=({\omega}_c^2-\omega^2)/\omega^2$, $M_{ck}=-\langle
f_c|V_c|f_k\rangle$, and $M_{kk}=(\omega_{k}^2-\omega^2)/\omega^2$. The
shorthand notation $V_c$, represents the perturbation in
the dielectric constant that results from adding in the cavity,
else there is a perfect PC (mirror) for $x>x_0$. Without the cavity, then 
$V=V_w$.

After solving the equation set by
matrix inversion,
the Green function is obtained analytically for the  complete waveguide-cavity system.
One obtains
\rev{
\begin{eqnarray}
\mathbf G^{\rm T}_{wc}\left|_{x>0 \atop x'>0} \right. &=&\mathbf G^{\rm
T}_{w}+\frac{\omega^2\ket{{\mathbf{f}}_c}\bra{{\mathbf{f}}_c}}{{\omega_c}^2-\omega^2-\omega^2\bra
{{\mathbf{f}}}V_c \mathbf G_{w}^{\rm T}V_c\ket{{\mathbf{f}}_c} }\nonumber\\
&+& \frac{\omega^2\mathbf G^{\rm
T}_{w}V_c\ket{{\mathbf{f}}_c}\bra{{\mathbf{f}}_c}}{{\omega_c}^2-\omega^2-\omega^2\bra
{{\mathbf{f}}_c}V_c \mathbf G_{w}^{\rm T}V_c\ket{{\mathbf{f}}_c} }\nonumber\\
&+& \frac{\omega^2\ket{{\mathbf{f}}_c}\bra{{\mathbf{f}}_c}V_c\mathbf
G^{\rm T}_{w}}{{\omega_c}^2-\omega^2-\omega^2\bra {{\mathbf{f}}_c}V_c
\mathbf G_{w}^{\rm T}V_c\ket{{\mathbf{f}}_c} }\nonumber\\
&+& \frac{\omega^2\mathbf G^{\rm
T}_{w}V_c\ket{{\mathbf{f}}_c}\bra{\tilde{\mathbf{f}}_c}V_c\mathbf G^{\rm
T}_{w}}{{\omega}_c^2-\omega^2-\omega^2\bra {{\mathbf{f}}_c}V_c \mathbf
G_{w}^{\rm T}V_c\ket{{\mathbf{f}}_c} },
\end{eqnarray}
 where $\mathbf G^{\rm T}_{wc}$ is in operator form,
 and by spatial projection:
 $\mathbf G^{\rm T}_{wc}(\mathbf r, \mathbf r')=\bra{{\bf r}}\mathbf G^{\rm T}_{wc}\ket{{\bf r}'}$, ${\mathbf f}_c (\mathbf r)=\langle \mathbf r\ket{{\mathbf f}_c}$.
}
Thus, the components of  $\mathbf G^{\rm T}_{wc}(\mathbf r, \mathbf r')$ projected onto ${\mathbf{f}_c(\mathbf r)}\otimes{\mathbf{f}^*_c(\mathbf r')}$,  and onto \rev{${\mathbf{f}_{k}(\mathbf r)}\otimes{\mathbf{f}_c^*(\mathbf r')}$},
are
%
\begin{eqnarray}
\mathbf G^{\rm T}_{cc}({\bf r},{\bf r}';\omega)\left|_{x>0 \atop x'>x_0} \right.  \!=\! \frac{\omega^2 {\mathbf{f}_c(\mathbf r)}\otimes{\mathbf{f}_c^*(\mathbf r')}}{{\omega}_c^2-\omega^2-i\omega(\Gamma_c^0+\Gamma_{wc}) } \ ,  \ \label{eq:3}
\end{eqnarray}
and
\begin{eqnarray}
\mathbf G^{\rm T}_{kc}({\bf r},{\bf r}';\omega)\left|_{0<x<x_0 \atop x'>x_0} \right.  \!=\! \frac{ia}{2v_g} \frac{A_{\rm fs}\, \omega^3 V_{kc} {\mathbf{e}_{k}(\mathbf r)}e^{ikx}\otimes{\mathbf{f}_c^*(\mathbf r')}}{ \omega_c^2-\omega^2-i\omega(\Gamma^0_c+\Gamma_{wc})}  \, , \ \ \ \  \label{eq:4}
\end{eqnarray}
where 
 $\Gamma^0_{c}$ is the vertical decay rate of the cavity, and $\Gamma_{wc}=A_{\rm fs}\,\Gamma_{wc}^0$ is the coupling  coefficient between the finite-size  waveguide and the cavity, with
 $A_{\rm fs}(L_{\rm eff},\omega)=1/[1+r^2-2r\cos(2k_\omega L_{\rm eff})]$
   and $\Gamma_{wc}^0=\frac{a\omega^2}{v_g}|V_{kc}|^2$; the latter term arises
   because of the evanescent coupling between the cavity mode and the waveguide mode,
   where
   $|V_{kc}|^2=|V_{ck}|^2 \approx| \int d\mathbf r \, \mathbf f_c^*(\mathbf r) V_w ({\bf r})
 {\bf e}_{k}(\mathbf r)e^{ikx}  |^2$. 
 In practical calculations, and in what follows below, we
 will compute this coupling exactly using a straightforward numerical simulation.
 Comparing with the
side-coupling waveguide-cavity system~\cite{hughes2004}, we highlight
 two important differences: the expression for  $\Gamma_{wc}^0$ is {\em doubled} with {\em unidirectional} coupling
 (for a side-coupled cavity, $\Gamma_{wc}^0=\frac{a\omega^2}{2v_g}|V_{kc}|^2$), and  there
  is a finite-size dependent coupling factor $A_{\rm fs}$.

The  Green function that describes propagation from the
 dot to the output waveguide can again be calculated by mode coupling theory, yielding
\begin{eqnarray}
\mathbf G^{\rm T}_{kc}({\bf r},{\bf r}';\omega)\left|_{x\ll 0 \atop x'>0} \right. \! =\!\frac{iat}{2v_g} \frac{A_{\rm fs}\, \omega^3 V_{kc} {\mathbf{f}_{k}^{\rm o}(y,z)}e^{ik^{\rm o}x}\otimes{\mathbf{f}_c^*(\mathbf r')}}{ \omega_c^2-\omega^2-i\omega(\Gamma^0_c+\Gamma_{wc})}  \, . \ \ \ \  \label{eq:5}
\end{eqnarray}


\rev{
\subsection{Enhanced Spontaneous Emission Regime}}
We can invoke the electric-dipole approximation
to derive the medium-dependent
spontaneous emission rate, or
 ({\em Einstein A coefficient}), defined through
\begin{eqnarray}
\Gamma({\mathbf r}_d, \omega_d)=\frac{2\mathbf{d}\cdot \rm{Im}[\mathbf
G^{\rm T}({\mathbf r}_d, {\mathbf r}_d;\omega_d)] \cdot\mathbf{d}}{\hbar
\varepsilon_0},
\end{eqnarray}
where \rev{$\mathbf d=\mathbf n_d d$} is the  optical dipole moment of the photon emitter's electronic resonance,
and ${\bf r}_d$ is the spatial position of the QD.
Therefore, the enhancement of spontaneous emission rate, i.e., the
 Purcell factor,  can be expressed analytically via  $F={\Gamma }/{\Gamma _{h}}$, where $\Gamma _{h}$ is the spontaneous
emission rate in
a corresponding homogeneous medium.
It is noted that
the concept of  spontaneous emission {\em rate} only makes sense for \rev{weak} and intermediate coupling regime.
  \rev{In other words the application of {\em Fermi's Golden Rule} assumes the weak coupling regime,
  which is an assumption that must be used with care for this system. However, our formalism is not
  restricted to this regime, and strong coupling effects
  will also be investigated later.}
Using the derived Green functions~(\ref{eq:3}-\ref{eq:4}), then
the on-resonance Purcell factor,
\rev{
\begin{equation}
F(L_{\rm eff},\omega_d=\omega_c)
=\frac{\Gamma_{\rm PC}}{\Gamma_{h}}
= \frac{6\pi c^3 |\mathbf n_d\cdot{\bf f}_{c}({\bf r}_d)|^2}{w^2\sqrt{\varepsilon_b}(\Gamma_c^0+A_{fs}\Gamma_{wc}^0)} ,
\end{equation}
 and the on-resonance beta factor,
 \begin{eqnarray}
  &&\beta (L_{\rm eff},\omega_d=\omega_c)=\frac{\Gamma_{\rm target}}{\Gamma_{\rm target} + \Gamma_{\rm others}},
  \nonumber\\
 &&
 \!\!\!\!\!\!\!=\!\frac{\int_{s_{\rm o}}{ \!\!{\rm Re}\{(\mathbf G^{\rm T}_{kc}({\bf r},{\bf r}_d;\omega_c)\!\cdot\! \mathbf n_d ) \!\!\times \!\! [{\bm \nabla\!}\!\!\times \!\!(\mathbf G^{\rm T}_{kc}({\bf r},{\bf r}_d;\omega_c)\!\cdot \!\mathbf n_d)} \! ]^*\} \cdot \!d\mathbf s }{\int_{s_{\rm d}} \!\!{ \!{\rm Re} \{ (\mathbf G^{\rm T}_{cc}({\bf r},{\bf r}_d;\omega_c)\!\cdot \!\mathbf n_d )\!\!\times  \!\![{\bm \nabla\!} \!\!\times  \!\! (\mathbf G^{\rm T}_{kc}({\bf r},{\bf r}_d;\omega_c)\!\cdot \!\mathbf n_d) ]^*} \}  \!\cdot \! d\mathbf s }\nonumber\\
 &&
 \ \ \ \ \ \ \ \ \ \ \ \ \ \ \ =  A_{fs}(L_{\rm eff},\omega_c) B_{\rm coup} ,
 \end{eqnarray}
where $B_{\rm coup}$ depends on the coupling out to the target waveguide, and is determined from the
full numerical simulation of a dipole, including the coupler region; $s_{\rm o}$
and $s_{\rm d}$ refer to surface regions perpendicular to the output
propagating waveguide (at $x\ll 0$) and to a surface surrounding the dipole, respectively.
The target output mode represents the output waveguide, and
we have neglected the influence of non-radiative decay since we are considering
QD coupling regimes at low temperature in an enhanced emission regime.
 These
analytical formulas are
valid for well defined PC waveguides, and, as we will show below, can even be used to accurately describe emission for integrated systems with only four unit cells in the waveguide section.}\\

\rev{
\subsection{Emitted Spectra and Strong Coupling Regime}
Assuming an {\em incoherently} excited QD, the exact electric field
operator can be written as~\cite{shughes08}
\begin{eqnarray}
\hat{\mathbf{E}}(\mathbf  R, \omega)&=&\frac{1}{\varepsilon_0}\mathbf{G}^{\rm T}(\mathbf R, \mathbf r_d;\omega)\cdot {\bf d} [\hat{\sigma}^-(\omega)+\hat{\sigma}^+(\omega)],
\end{eqnarray}
where  $\hat\sigma^{(\pm)}$ are the Pauli operators of the
electron-hole  pair (exciton).
The spectrum, detected at position ${\bf R}$, is~\cite{shughes08}:
\begin{eqnarray}
S({\bf R},\omega) &  = & |{\bf G}^{\rm T}({\bf R},{\bf r}_d;\omega) \cdot {\bf d}|^2 \times \nonumber \\
&\hbox{} &
\!\!\!\!\!\!\!\!\!\!\!\!\!\!\!\!\!\!\!\!\!\!\!\!\!\!\!\!\!
\left |\frac{\alpha_0(\omega)}
{(1-\alpha_0(\omega)\, {\bf n}_d \cdot
{\bf G}^{\rm T}_{cc}({\bf r}_d,{\bf r}_d;\omega) \cdot {\bf n}_d )} \right|^2\!\!, \ \ \ \ \
\end{eqnarray}
where $\alpha_0(\omega)=2\omega_d d^2/[\hbar\varepsilon_0(\omega_d^2-\omega^2)]$ is the bare
polarizability, with $\omega_d$ the exciton resonance frequency. It is noted that the contribution from continuous radiation modes have been neglected, which cause the divergence of Green function at $\mathbf r=\mathbf r'$ and the vacuum Lamb shift; since
this shift is typically very small and can be thought to exist already
in the definition of $\omega_d$, it can be safely neglected.
Using ${\bf G}^{\rm T}_{cc}({\bf R},{\bf r}_d;\omega)$ and ${\bf G}^{\rm T}_{kc}({\bf R},{\bf r}_d;\omega)$
from Eqs.~(\ref{eq:3}-\ref{eq:4}), we obtain the spectrum at any relevant spatial point, e.g.,
above the cavity, or along the
output  waveguide. For example, when the photon is emitted
on-chip along the waveguide, then
\begin{eqnarray}
S_{\rm side}({\bf R},\omega) &  \approx & |{\bf G}_{kc}^{\rm T}({\bf R},{\bf r}_d;\omega) \cdot {\bf d}|^2 \times \nonumber \\
&\hbox{} &
\!\!\!\!\!\!\!\!\!\!\!\!\!\!\!\!\!\!\!\!\!\!\!\!\!\!\!\!\!
\left |\frac{\alpha_0(\omega)}
{(1-\alpha_0(\omega)\, {\bf n}_d \cdot
{\bf G}^{\rm T}_{cc}({\bf r}_d,{\bf r}_d;\omega) \cdot {\bf n}_d } \right|^2\!\!, \ \ \ \ \
\end{eqnarray}
and when the photon is emitted vertically, above the cavity:
\begin{eqnarray}
S_{\rm vert}({\bf R},\omega) &  \approx & |{\bf G}_{cc}^{\rm T}({\bf R},{\bf r}_d;\omega) \cdot {\bf d}|^2 \times \nonumber \\
&\hbox{} &
\!\!\!\!\!\!\!\!\!\!\!\!\!\!\!\!\!\!\!\!\!\!\!\!\!\!\!\!\!
\left |\frac{\alpha_0(\omega)}
{(1-\alpha_0(\omega)\, {\bf n}_d \cdot
{\bf G}^{\rm T}_{cc}({\bf r}_d,{\bf r}_d;\omega) \cdot {\bf n}_d  } \right|^2\!\!. \ \ \ \ \
\end{eqnarray}}
We now have all the relevant formulas to compute the Purcell factor, beta factor,
and emission spectrum for the integrated waveguide-cavity system shown in Fig.~(1).
}


\rev{
\section{Calculations}}

\rev{
\subsection{Weak Coupling Regime}
}

 In order to validate the above Green function theory, a
 direct 3D finite-difference time-domain (FDTD)  calculation of the Green function
 terms is first performed~\cite{YaoLPR2009,Lum}.
 We use parameters
  representative of the popular L3 cavity~\cite{noda}  and
a  nominal W1 (removed row of holes) waveguide, with the following parameters: semiconductor slab dielectric
 constant $\varepsilon=12$; lattice constant $a=420\,$nm (PC pitch);
 the two holes as indicated in Fig.~1 are shifted outwards by a  distance of $0.15\,a$; the thickness of the slab and radius $R$ of the air holes are $0.5\,a$ and $0.275\,a$, respectively; the width
of the output waveguide is  470\,nm, which was optimized to give the largest
beta factor.
The TE-like band gap ranges from  0.760\,eV to 0.935\,eV (corresponding to 0.26--0.32
$c/a$ in normalized frequency units, or 85 to 228\,THz), and the
band structure of the waveguide mode is shown in Fig.~2(a). In the frequency range of our interest, the waveguide is single mode and under the light line (gray shaded region). In Fig.~2(b), we show the  enhancement of the spontaneous emission  versus frequency for  a maximally positioned and $y-$aligned QD exciton, with $L=6a$;
  the Purcell factor spectra  exhibits a typical Lorentzian  line shape, that agrees with
the analytical expression of Eq.~(1).  The electric-field distribution
at the resonant frequency (indicated by red circle in Fig.~2) is also shown in  Fig.~3.
  The local field strength in the cavity is
pronounced
and the energy is mainly guided
into the coupled PC waveguide, and subsequently into the target output
waveguide;
\rev{both a significant Purcell factor and an enhanced beta factor
are obtained.
Although the Purcell factor is reduced in comparison to a bare waveguide, the emphasis
here is on achieving an enhanced Purcell factor while still obtaining
a large on-chip $\beta-$factor.
}
These Purcell factors give a quantitative measure
of the
enhancement in the projected LDOS, and are already large enough,
with suitable
QD coupling, to facilitate strong coupling.
The effective mode volume of the cavity system is found
to be $V_{\rm eff}\approx 0.063\,\mu{\rm m}^3$,
which can be related to the cavity mode position
at the peak field antinode, through
$|{\bf f}_c({\bf r}_{\rm antinode})|^2=1/V_{\rm eff}\varepsilon$.

\begin{figure}[t]
\centering
\includegraphics[width=0.46\textwidth]{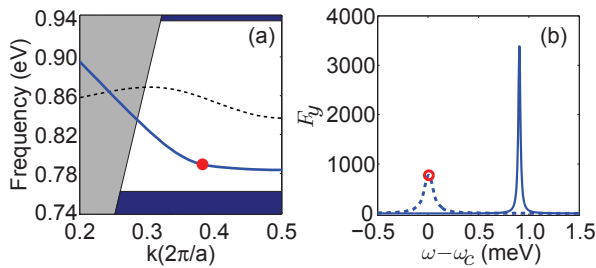}
\caption{(color online) (a) The TE-like band structure of the planar PC  (W1) waveguide (see Fig.\,1). The filled red dot indicating the waveguide-cavity resonant frequency  $\omega_c=192.54$\,THz, with a  corresponding $k({\omega_c})=0.75\pi/a$. (b) Theoretical
maximum Purcell factors of cavity-waveguide system versus frequency when $L=6a$
(dashed curve),
the resonant frequency is labeled by red circle. \rev{For reference,
we also show the bare cavity results with the solid curve, for a cavity surrounded by a large number of holes on all sides; the frequency shift is a result of the different boundary condition for the finite-size cavity.}}
\end{figure}
\begin{figure}[t]
\hspace{0cm}
\centering
\includegraphics[width=0.44\textwidth]{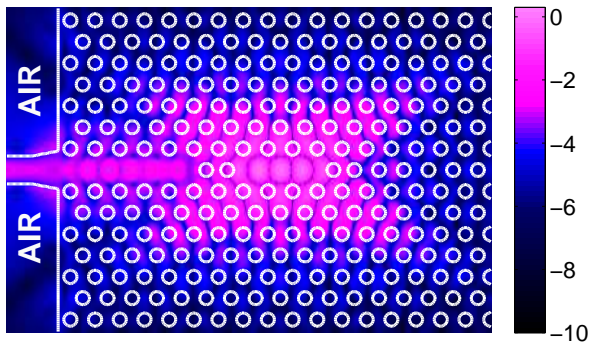}
\caption{\label{fig:cavit_with_waveguide} (color online) The  distribution of
electric field amplitude ($|{\bf E}(\omega_c)|$) at slab center plane on a log scale,
where $\omega_c$  is the peak Purcell factor frequency shown in
Figure 3(b).}
\end{figure}


\begin{figure}[t]
\vspace{-0.50cm}
\centering
\includegraphics[width=0.46\textwidth]{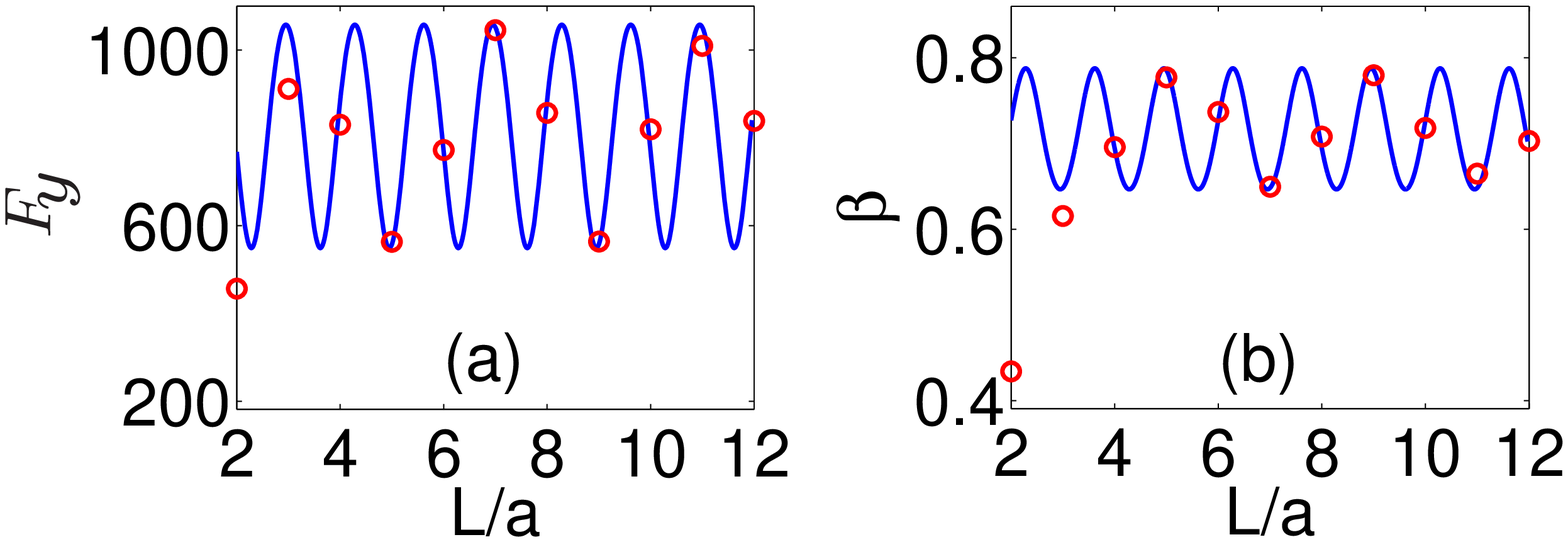}
\vspace{-0.3cm}
\caption{\label{fig:cavit_with_waveguide} (color online) The dependence of (a) Purcell factor and (b)
beta factor
as a function of waveguide length,  $L$.  The data indicated by red circles is obtained
 from the full 3D numerical simulation, while  the blue curves show the results from
 the derived analytical expression.}
 \vspace{0.30cm}
\end{figure}

Next,  we carry out
  a systematic investigation of the Purcell factor  as a function of  length $L$,
    where $L$ is increased by integer multiple of $a$.
The results are shown in Fig.\,4(a), and the data is successfully fitted with  the
analytical form introduced earlier;
the main parameters required for the analytical formulas are extracted
from carrying out  one numerical simulation,
to obtain
 $r=0.21$ and $\Gamma^0_{c}/\Gamma_{wc}^0=0.31$. The center wave vector $k({\omega_c})=0.75\pi/a$
 is obtained
 from the band structure.
  From Fig.~5, we know when $L$ is larger than $3a$,
  then \rev{the  Bloch mode theory becomes valid}. \rev{Of course, the analytical theory fit is only effective for integer multiple of $a$ because of the coupler-termination dependence of $r$. If we want to show the case of a continuously varying $L$, we should first  calculate $r$ (and $t$) for various unit cell truncations at the output coupler. However,
  since we have optimized this coupler region, the most practical case is for the integer number
  of unit cells.}
 Importantly, our calculations  include the output reflection coefficient and the length of waveguide.
  In addition, one can also tune the properties of the cavity, e.g., to the
  target exciton resonance, and still overlap with the
  broadband coupling region of the PC waveguide mode (20-40\,meV bandwidth
  below the light line, cf. Fig.~2(a)).

We also investigate the single-photon  $\beta-$factor; this parameter
 quantifies the efficiency of emitting a single photon into the
 desired output mode, namely
 the non-PC waveguide mode after the coupler (cf.~Figs.~1 and 2).
The beta factor is first calculated by using the
 numerically-exact FDTD technique, by
computing the emitted fields at the left of the coupler; these
fields  are
subsequently
mode-overlapped with the desired waveguide mode and normalized with respect
to the total power flowing out of the lossless device.
\rev{We first obtain the total emitted power $P_t$ by having six field monitors completely surrounding the emitting dipole; we also record the propagation power after the field travels through the coupler $P_{\rm out}$,  and
 calculate the field distribution $\mathbf E_i(\mathbf r)$ and  $\mathbf H_i(\mathbf r)$,
  including all bound and radiation mode contributions.
  We then adopt a mode overlap integral technique,
 and calculate the projection of the scattered field that overlaps with the target
 output waveguide mode.
   Labeling the electric and magnetic field of the target mode as $\mathbf E_{\rm out}(\mathbf r)$ and
   $\mathbf H_{\rm out}(\mathbf r)$,
   respectively, then the overlap integral can be expressed as
\begin{eqnarray}
OI&=&\frac{{\rm Re}\left[\frac{\int \mathbf E_{\rm out}(\mathbf r)\times \mathbf H_i^*(\mathbf r))\cdot d\mathbf S\int \mathbf E_i(\mathbf r)\times \mathbf H_o^*(\mathbf r))\cdot d\mathbf S}{\int \mathbf E_{\rm out}(\mathbf r)\times \mathbf H_{\rm out}^*(\mathbf r))\cdot d\mathbf S}\right]}{{{\rm Re}(\int \mathbf E_i(\mathbf r)\times \mathbf H_i(\mathbf r))\cdot d\mathbf S}}, \
\end{eqnarray}
where $S$ is on the $y-z$ plane perpendicular to the target waveguide
at the left of the coupler region ($x\ll 0$). The beta factor is then simply $\beta=OI\times {P_o}/{P_t}$.} The key advantages in the present proposal are as follows: $i$) \rev{the  Purcell factors, in comparison to a  bare finite-size waveguide, are substantially higher
 when over-coupled to the cavity},
 $ii)$ the
 Purcell factor and beta factor can be controlled in a systematic way,
 and $iii)$, the conceptional understanding
 and coupling can be described analytically; to show that this latter point
 is also true for the beta factor, we have
 fitted the analytical form with the previous parameters
 and found
 good agreement when $L\geq 4\,a$ (Fig.~4(b)).

%
\begin{figure}[t]
\hspace{0cm}
\centering
\includegraphics[width=0.44\textwidth]{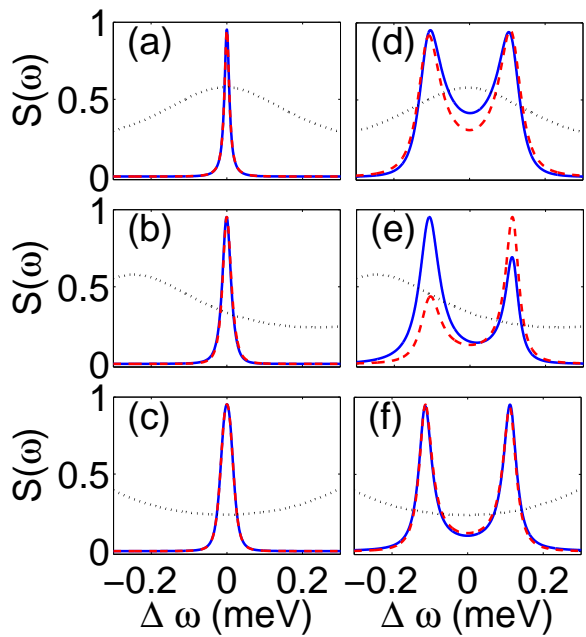}
\vspace{-0.1cm}
\caption{
(color online) Spectrum emitted vertically from the cavity mode
(blue solid curve) and on-chip via the output waveguide mode (red dashed curve); also shown is
the finite-size coupling coefficient $A_{\rm fs}$ (black dotted curve).
(a-c) $d=30$\,Debye   for $L=9\,a,10\,a,11\,a$, respectively.
(d-f) $d=50$\,Debye and $r=0.21$, for $L=109\,a,110\,a,111\,a$, respectively.}
\end{figure}


\rev{
\subsection{Strong Coupling Regime}}
\rev{Finally, we turn our attention to the strong coupling regime.}
Since the coupled QD-PC system  results in significant LDOS
enhancements, we can naturally probe strong coupling and non-perturbative
cavity-QED, either above the cavity or along the output waveguide.

\rev{We study emission spectra for detectors that are placed above the cavity and
at the target waveguide, for various waveguide lengths. These represent the
spectra emitted off-chip and on-chip. For these strong coupling calculations, one must include
the dispersion of the propagating PC waveguide mode; using a linear dispersion model, 
$k_\omega = k_{\omega_d} + (\omega-\omega_d)/v_g$, where $v_g \approx 14$ is obtained
from the slope of the waveguide band in Fig.~3(a), at the indicated red circle.}

In Fig.~6(a-c), we
display the emitted spectra, both vertically and for the output waveguide,
for
a $9-11\,a$ unit-cell PC waveguide, and
a dipole moment of $d=30$ Debye; to ensure maximum coupling,
the exciton is resonant with the  cavity mode ($\omega_d=\omega_c$).
Clearly the emitted spectra are
qualitatively different depending upon the PC waveguide length, which is
 due to the modal propagation characteristics
of the cavity and the finite size waveguide; \rev{in particular, one can see the broadening and
thus the Purcell factor increases as we go from
   ($L=9\,a$) to  ($L=11\,a$), showing that
   the QD coupling depends sensitively upon the
 length of the waveguide section.
  Ideally, for a side-coupled waveguide-cavity system, with no external reflection from the
   waveguide ends, the shape of spectra emitted vertically from cavity mode and on-chip via the waveguide mode are symmetric  and  identical. However, for any real system, the effect of finite size is always there,
    and in general will result in different spectral shapes for the
    vertically and horizontally emitted spectra, e.g., proportional to $A_{\rm fs}(L_{\rm eff},\omega)$.}

  We next choose a  slightly larger dipole moment of $50\,$Debye and a  longer PC guide, with $L=109-111\,a$ unit-cells.
  \rev{A PC waveguide length of $109-111a$ has the same peak PF as those with $9-11a$, since
  the length difference between them ($100a$)  is an integer multiple of the period, $8a/3$ ($k_{\omega_d}=0.75\pi/a$); at the resonance frequency, the Purcell factor is periodic, however, for off-resonance, it will be more complicated because the field is propagating back and forth between the coupler and the cavity, which appears differently
  in general for on-chip emission and out-of-plane (vertical) emission
  (cf.~the presented spectrum equations (16) and (17)).}
As shown in  Fig.~6(d-f), we recognize a much
  larger frequency-dependence on the coupling parameter
  $A_{\rm fs}(L_{\rm eff},\omega)$, which can produce significant asymmetries
  in the emission spectra of a single QD exciton;
indeed,  there is now a substantial difference between the vertically emitted
  light and the emitted light on chip, and, in principle, these different
  spectra could be probed in experiments by placing detectors above the cavity
  and at the output of the exit waveguide.\\



\rev{
\section{Conclusions}
}
 We have proposed
and investigated
 the spontaneous emission properties of an embedded single  QD in a
 photonic crystal waveguide-cavity system.
To describe the quantum light-matter interactions in this system,
an intuitive
Green function
formalism has been developed which
is confirmed by  detailed numerical calculations.
  The structure can achieve both large Purcell
 factors and high extraction rates, and
allow the investigation of weak and strong coupling regimes, both on- and
off-chip.\rev{ These waveguide-cavity
 systems are timely with recent improvements in PC fabrication, and offer a
very
   rich degree of
  fundamental control of the
  ensuing light-matter interactions.}\\


\rev{
\section*{Acknowledgments}
}
This work was supported by the National Sciences and Engineering Research Council of Canada,
and the Canadian Foundation for Innovation. We thank Mark Patterson for assistance in
carrying out the PC band structure calculations.




\begin{thebibliography}{1}


\rev{
\bibitem{moreau} E. Moreau,
I. Robert, J. M. G\'{e}rard, I. Abram, L. Manin, and V. Thierry-Mieg,
Appl. Phys. Lett.
\textbf{79}, 2865 (2001).

\bibitem{santori2} C. Santori,
D. Fattal, J. Vu{\u{c}}kovi\'{c},  G. S. Solomon,  and  Y. Yamamoto,
Nature \textbf{419},  594 (2002).

\bibitem{pelton} M. Pelton, C. Santori, J. Vu{\u{c}}kovi\'{c}, B. Zhang, G. S. Solomon, J. Plant, and Y. Yamamoto,
Phys. Rev. Lett. \textbf{89}, 233602 (2002).


\bibitem{hennessy} K. Hennessy,
A. Badolato, M. Winger, D. Gerace, M. Atat\"ure, S. Gulde, S. F\"alt, E. L. Hu, and  A. Imamo\u glu,  Nature \textbf{445}, 896 (2007).

\bibitem{englund} D. Englund, D. Fattal, E. Waks, G. Solomon, B. Zhang, T. Nakaoka,
Y. Arakawa, Y. Yamamoto, and J. Vu{\u{c}}kovi\'{c},
Phys. Rev. Lett. \textbf{95}, 013904 (2005).



\bibitem{kuhnPRB02}
see, e.g., A. Vagov, V. M. Axt, and T. Kuhn,
Phys. Rev. B {\bf 66}, 165312 (2002).





\bibitem{purcell} E. M. Purcell,
Phys. Rev. \textbf{69}, 681(1946).


\bibitem{noda} Y. Akahane,
T. Asano, B. Song, and S. Noda,
Nature \textbf{425}, 944(2003).


\bibitem{badolato}A. Badolato,K. Hennessy, M. Atat\"ure,
J. Dreiser, E. Hu, P. M. Petroff, A. Imamo\u glu,
 Science \textbf{308}, 1158(2005).%







\bibitem{thorhauge}M. Thorhauge, L. H. Frandsen, and P. I. Borel,
Opt. Lett. \textbf{28},
1525 (2003).

\bibitem{fan} S. Fan, P. R. Villeneuve, J. D. Joannopoulos, and H. A. Haus,
Phys. Rev. Lett. \textbf{80},  960 (1998).

\bibitem{wu} L. Wu, M. Mazilu, J. -. Gallet, T. F. Krauss, A. Jugessur, and
R. M. De La Rue,
Opt. Lett. \textbf{29}, 1620 (2004).


\bibitem{jelOL2007}
D. Englund, Andrei Faraon, Bingyang Zhang, Yoshihisa Yamamoto, and Jelena Vuckovic,
Opt. Express {\bf 15},  5550 (2007).

\bibitem{YaoPRL2005}
W. Yao, R-B Liu, and L. J. Sham, Phys. Rev. Lett. 95, 030504 (2005)

\bibitem{GaoAPL2008} Jie Gao, F.W.Sun, and Chee Wei Wong,
Appl. Phys. Lett. {\bf 93}, 151108 (2008).


\bibitem{YangPRL2009}
Xiaodong Yang, Mingbin Yu, Dim-Lee Kwong, and Chee Wei Wong,
 Phys. Rev. Lett.  {\bf 102}, 173902 (2009).



\bibitem{lecamp}  G. Lecamp, P. Lalanne, and J. P. Hugonin,
Phys. Rev. Lett. \textbf{99}, 023902 (2007).



\bibitem{hughes2004} S. Hughes,   Opt. Lett. \textbf{29}, 2659 (2004).


\bibitem{hughesprb} V. S. C. Manga Rao and S. Hughes, Phys. Rev. B \textbf{75}, 205437 (2007).


\bibitem{viasnoffschwoob} E. Viasnoff-Schwoob, C. Weisbuch, H. Benisty, S. Olivier, S. Varoutsis, I. Robert-Philip,
R. Houdr\'e, and C. J. M. Smith,
Phys. Rev. Lett.
\textbf{95}, 183901 (2005).





%
\bibitem{lund}T. Lund-Hansen, S. Stobbe, B. Julsgaard, H. Thyrrestrup,
T. S\"unner, M. Kamp, A. Forchel, and P. Lodahl, 
Phys. Rev. Lett. \textbf{101}, 113903 (2008).



\bibitem{Hughes:2005} S. Hughes, L. Ramunno, J. F. Young, and J. E. Sipe,
{{Phys. Rev.
Lett.}} \textbf{{94}}, {033903} ({2005}).


\bibitem{Povinelli:2004}
M. L. Povinelli,   S. G. Johnson, E. Lidorikis, J. D.
Joannopoulos, and Marin Solja\v{c}i\'{c}, 
{App. Phys. Lett.}
  \textbf{{84}}, {3639} ({2004}).



\bibitem{Gerace:2004} D. Gerace, and L. C. Andreani,
defects, Opt. Lett. \textbf{29}, 1897 (2004).




\bibitem{Fussell:2008} D. P. Fussell,
S. Hughes, and M. M. Dignam,
Phys. Rev. B \textbf{78}, 144201 (2008).



\bibitem{manga_photongun} V.S.C. Manga Rao and S. Hughes, Phys. Rev. Lett. \textbf{99}, 193901 (2007).



\bibitem{Banaee:APL2007}
M.G. Banaee, A.G. Pattantyus-Abraham, M.W. McCutcheon, G.W. Rieger, and J.F. Young,
Appl. Phys. Lett.
\textbf{90}, 193106 (2007).

\bibitem{young04}
A. Cowan and J. E. Young,  Phys. Rev. E
{\bf 68}, 046606 (2003).




\bibitem{shughesprb04} S. Hughes and H.
Kamada,   Phys.
Rev. B. {\bf 70}, 195313 (2004).

\bibitem{YaoLPR2009}
P. Yao, V.S.C. Manga Rao, and S. Hughes,
Laser and Photonics Review, DOI 10.1002/lpor.200810081 (2009).

\bibitem{Lum} For our FDTD calculations, we have used
Lumerical Solition Inc.: see www.lumerical.com





\bibitem{phase} Alternatively,
one could also define the phase $(2kL+\phi)$,
instead of $(2kL_{\rm eff})$, so
$L_{\rm eff}$ incorporates $\phi$.






\bibitem{shughes08}
S. Hughes and P. Yao,
{Opt. Express \textbf{17}, 3322 (2009).}
}

\end{thebibliography}
\end{document}